\documentclass[sigconf]{acmart}

\AtBeginDocument{%
  \providecommand\BibTeX{{%
    \normalfont B\kern-0.5em{\scshape i\kern-0.25em b}\kern-0.8em\TeX}}}

\setcopyright{acmcopyright}
\copyrightyear{2022}
\acmYear{2022}
\acmDOI{XXXXXXX.XXXXXXX}

\acmConference[RoPES'22]{Co-Located with ICSE 2022}{1st International Workshop on Recruiting Participants for Empirical Software Engineering}{Pittsburg, USA}

\acmSubmissionID{123-A56-BU3}

\usepackage[htt]{hyphenat}
\usepackage{seqsplit}
\usepackage[vskip=0.5em,font=itshape,leftmargin=2em,rightmargin=0.5em]{quoting}

\usepackage{url}
\usepackage{cmap}

\begin{document}

\title[Privacy Perils of MSR for Participants]{Should I Get Involved? On the Privacy Perils of Mining Software Repositories for Research Participants}

\author{Melina Vidoni}
\email{melina.vidoni@anu.edu.au}
\orcid{0000-0002-4099-1430}
\affiliation{%
  \institution{Australian National University}
  \country{Australia}
}

\author{Nicolás E. Díaz Ferreyra}
\email{nicolas.diaz-ferreyra@tuhh.de}
\orcid{0000-0001-6304-771X}
\affiliation{%
  \institution{Hamburg University of Technology}
  \country{Germany}
}

\renewcommand{\shortauthors}{Vidoni and Díaz Ferreyra}

\begin{abstract}
Mining Software Repositories (MSRs) is an evidence-based methodology that cross-links data to uncover actionable information about software systems. Empirical studies in software engineering often leverage MSR techniques as they allow researchers to unveil issues and flaws in software development so as to analyse the different factors contributing to them. Hence, counting on fine-grained information about the repositories and sources being mined (e.g., server names, and contributors' identities) is essential for the reproducibility and transparency of MSR studies. However, this can also introduce threats to participants' privacy as their identities may be linked to flawed/sub-optimal programming practices (e.g., code smells, improper documentation), or vice-versa. Moreover, this can be extensible to close collaborators and community members resulting ``guilty by association''. This position paper aims to start a discussion about indirect participation in MSRs investigations, the dichotomy of `privacy vs. utility' regarding sharing non-aggregated data, and its effects on \textit{privacy restrictions} and \textit{ethical considerations for participant involvement}.
\end{abstract}

\begin{CCSXML}
<ccs2012>
   <concept>
       <concept_id>10002978.10003029.10011150</concept_id>
       <concept_desc>Security and privacy~Privacy protections</concept_desc>
       <concept_significance>500</concept_significance>
       </concept>
   <concept>
       <concept_id>10002978.10002991.10002995</concept_id>
       <concept_desc>Security and privacy~Privacy-preserving protocols</concept_desc>
       <concept_significance>500</concept_significance>
       </concept>
   <concept>
       <concept_id>10011007.10011006.10011072</concept_id>
       <concept_desc>Software and its engineering~Software libraries and repositories</concept_desc>
       <concept_significance>500</concept_significance>
       </concept>
 </ccs2012>
\end{CCSXML}

\ccsdesc[500]{Security and privacy~Privacy protections}
\ccsdesc[500]{Security and privacy~Privacy-preserving protocols}
\ccsdesc[500]{Software and its engineering~Software libraries and repositories}

\keywords{Mining software repositories, developers' privacy, ethical considerations, privacy-preserving protocols, indirect participants}

\maketitle

\section{Position Motivation}

Let us consider the following motivational example:

\textbf{Motivational Example}. A group of researchers mine several C++ projects from GitHub with the aim of spotting security flaws and determine gaps in developers' cybersecurity training. Following existing recommendations, they craft a detailed inclusion/exclusion criteria, perform a systematic repository search, and extract developers' emails from their associated GitHub's profiles. Using the latter, they distribute a survey to elicit their knowledge and practices regarding software security. The process undergoes an Ethical Protocol examination, which is approved, and requests participants' data (e.g., emails, names) to be anonymised.% (i.e., emails and names will not be shared). %The Protocol is approved, and the researchers proceed with the study.

\textbf{Problem Statement.} Several issues arise from the motivational example. Due to open science standards, it is increasingly common to provide replication packages, which may disclose the analysed sources by name, or through non-aggregated data (e.g., code snippets, function signatures). How this data is compiled and reported can be unclear for the study participants as this information is often absent in the Participant Information Sheet (PIS). Moreover, sometimes is not even explained to them nor addressed within the Ethical Protocol. By disclosing non-aggregated (sometimes \textit{raw}) data, researchers may allow participants' identities to be traced back, even for those who did not respond the survey. 

The latter can be considered `indirect participants', as in medical research: ``those whose rights and welfare may be affected by the intervention through their routine exposure to the environment in which the intervention is being deployed" \citep{Smalley2015}. Within the context of MSRs, the `intervention' can be interpreted as the analyses or studies conducted using the data being mined from the software repositories; disclosing the findings and the repositories by name (or with traceable data) allows the re-identification of `indirect participants'.

This situation can nonetheless lead to `guilty by association' issues. For instance, let us consider the case of a project A in which many developers collaborated but only a few of them contributed largely (the rest just had a secondary role). Imagine that one of these active members (who is very security unsavvy) takes the survey described in the motivational example. Now imagine that, after analysing the mined repositories and processing the survey results, the researchers conclude that security flaws in open-source projects are closely related to gaps in the cybersecurity training of their contributors. If the researchers disclose project A's name alongside the identified security flaws, one could think that all developers who were involved in the project are unsavvy when, in fact, only a bunch of them are responsible for the security flaws and the rest are just `guilty by association'.

%, in a worst-case-scenario, recruiters could extrapolate that C1 and C2 have a deficient training. 

Results may negatively affect practitioners' careers if expose (as in the example above) less-than-ideal practices (e.g., anti-patterns, security breaches) along with attributes allowing participant's re-identification. Data linkage is a well-known privacy threat in the Internet of Things as general information about a community or its members can be easily inferred from the data collected by different devices \citep{Zheng2018}. This poses an issue regarding the \textit{ethical considerations of participant involvement} (e.g., informed consent) and whether or not the research may negatively impact their careers. Information leakage in data mining and its potential privacy breaches is a known risk \citep{PRAKASH2015}, which has reportedly compromised participants' privacy through indirect data sharing \citep{Pournajaf2016}. 

On the other hand, presenting just aggregated data (to prevent these issues or due to the Ethical Protocol) may limit the study's reproducibility and lead to peer-review criticism. However, providing non-aggregated/raw data may also affect \textit{privacy restrictions} (e.g., anonymity and confidentiality) or even bypass the data-protection boundaries enforced by \textit{ethical committees}. Although some privacy-preserving data-linkage protocols have been proposed within the current literature \citep{OKeefe2004, AlAzizy2014}, their expansion to MSRs and mixed-methods remains unclear. Similarly, the analysis of outliers in datasets may lead to the identification of private information regarding those particular cases, as most anonymisation techniques (both semantic and syntactic) do not consider outliers in their process \citep{WANG201594}.

\section{Perils of Involvement}

The problems derived from the motivational example may have unwanted and negative consequences. 

\textsc{Peril I.} The likelihood of data being retraced and identities uncovered may hinder participants' willingness to partake, by (a) not returning surveys (affecting \textit{self-selection bias} and causing \textit{sample sizes issues} \citep{Baltes2020}), (b) providing false information in the survey, (c) requesting not to be contacted in the future (affecting future works), or (d) requesting their repository to be withdrawn.

\textsc{Peril II.} Data traceability may discourage developers from contributing to OSS (open-source software) repositories or persuade them to enact `barriers' to prevent indirect participation \citep{Smalley2015}; e.g., licences limiting which investigations can be conducted \citep{Gold2020}. This is an \textit{ethical consideration} regarding \textit{participant involvement}, because the consent over who can use community resources in OSS is crucial, reverting to several questions posed by prior research \citep{Berry2004}:

\begin{quoting}
Is it necessary for each individual in a community to give consent to research before research can be conducted, or can a few individuals represent the group? This raises two important distinctions: (1) how can a community give consent before they are researched, and (2) should a community be required to consent to the publication of the results of research.
\end{quoting}

\textsc{Peril III.} Researchers are the `guardians' of the participants' privacy, and may incur in indirect disclosure of personal data (related to \textit{privacy restrictions} and \textit{ethical considerations}). Current trends in open science endorse sharing a primary dataset alongside articles\footnote{See: \url{https://conf.researchr.org/track/icse-2022/icse-2022-open-science-policies}}. However, as \citet{Erb2021} stated, ``while an informed consent given by study participants may satisfy legal requirements, and the removal of obvious identifiers prevents trivial re-identification, we argue that this perspective is typically too short-sighted to address actual ramifications of public data releases in terms of anonymity". This topic is not new, and many disciplines have analysed it \cite{Thorogood2018, Colin2018}. Nevertheless, it remains foreign to MSRs studies and direct/indirect participants.

\section{The Path Forward}

Preserving participants' privacy while maximising the utility of their personal data is an old trade-off conundrum. \citep{Li2009}. Still, it has repercussions on individuals' privacy \cite{Erb2021} that go beyond the sinless use of contact details in MSR studies. The nuances of ethical mining and data disclosure \citep{Gold2020} can harm MSRs participants (both direct and indirect) while limiting the recruitment and engagement in future research. However, the impact of MSRs and open science on participants' privacy must be closely examined. Hence, this position proposes three areas for future works:

\textsc{Proposal I}. Other disciplines posed privacy-aware methodologies to prevent deanonymisation through data linkage \citep{OKeefe2004, AlAzizy2014}, and guidelines for handling data disclosure on research depending on information retrieval and participant recruitment \citep{Redmiles2017}. Investigating their transfer to mixed-method MSR studies is a future line of work, relevant to the participants' \textit{privacy restrictions} and the \textit{ethical considerations} of their involvement. This can provide guidelines for data sharing and presentation related to \textit{Peril III}.

\textsc{Proposal II}. Participant recruitment in mixed-method studies extends beyond contacting people to complete a survey, partake in workshops or interviews. Further assessing the nuances of participation, consent and involvement \citep{Berry2004}, may lead to the evolution of ethical mining approaches \citep{Gold2020} to mitigate the risk of retracing and deanonymising data \citep{Zheng2018, Pournajaf2016}. This involves assessing indirect participation in MSRs \citep{Smalley2015}, and issues posed in \textit{Perils I-II}.

\textsc{Proposal III.} Balancing the `privacy vs utility' dichotomy \citep{Li2009}, to investigate trends in MSRs studies and their effects on participants (both direct/indirect). Open science in software engineering, particular MSRs, needs to bring the discussion of participants' data-protection rights and information disclosure to the forefront \citep{Erb2021}. This would enable \textit{institutional review boards} to have guidelines for the assessment of Ethical Protocols and would assist the peer-review process of ethically-obscured methodologies. This relates to \textit{Perils II-III}.

\section{Conclusions}

This position aims to present the problem of OSS contributors as an \textit{ethical consideration of participant involvement} in MSR studies while discussing examples of existing situations that affect \textit{privacy restrictions}. The overall goal is to highlight the problems from the lens of the `privacy vs utility' spectrum and propose further lines of research that will contribute to the robustness and ethical conduction of MSRs (both pure and mixed-method) investigations.

Several lines of work are available as a path forward. First, to investigate what are the current perceptions of researchers regarding privacy considerations of MSR studies, and compare them to the `actual' privacy measures they follow (if any). Second, evaluating the `privacy vs utility' dichotomy in MSR papers, to produce guidelines that would help researchers decide when to prioritise reproducibility over privacy (and vice-versa).

\bibliographystyle{ACM-Reference-Format}
{\footnotesize \bibliography{references}}

\end{document}